\documentclass{appolb}
\usepackage{graphicx}        % standard LaTeX graphics tool
\usepackage{axodraw}
\usepackage{amssymb,wasysym}
\def\amuh{a_\mu^{{\mathrm{had}}}}

\newcommand{\mbo}[1]{$#1$}
\newcommand{\semis}{\;;\;\;}

\newcommand{\I}{{\rm i}}
\newcommand{\varv}{v}

\newcommand{\ppm}{\pi^+\pi^-}
\newcommand{\ba}{\begin{eqnarray*}}
\newcommand{\bea}{\begin{eqnarray}}
\newcommand{\Ba}{\begin{eqnarray}}
\newcommand{\beq}{\begin{equation}}
\newcommand{\bet}{\begin{center} \begin{tabular}}
\newcommand{\be}{\begin{eqnarray*}}
\newcommand{\bit}{\begin{itemize}}
\newcommand{\ea}{\end{eqnarray*}}

\newcommand{\epo}{\;\;. }
\newcommand{\eea}{\end{eqnarray}}
\newcommand{\Ea}{\end{eqnarray}}
\newcommand{\eeq}{\end{equation}}
\newcommand{\ee}{\end{eqnarray*}}
\newcommand{\eit}{\end{itemize}}

\newcommand{\ent}{\end{tabular} \end{center}}
\newcommand{\gv}{\mbox{GeV}}

\newcommand{\mv}{\mbox{MeV}}

\newcommand{\noi}{\noindent}

\newcommand{\cL}{{\cal L}}

\newcommand{\ttc}[1]{\multicolumn{2}{c}{#1}}

\newcommand{\epm}{e^+e^-}

\newcommand{\power}[1]{\cdot 10^{#1}}
\newcommand{\nc}{\newcommand}
\nc{\pa}{\partial}
\nc{\parsym} {\stackrel{\leftrightarrow}{\pa}}
\newcommand{\eepp}{e^+e^-\to \pi^+\pi^-}

\nc{\omg}{\omega}

\DeclareMathSymbol{\varPhi}{\mathalpha}{operators}{"08}
\DeclareMathSymbol{\varOmega}{\mathalpha}{operators}{"0A}

\begin{document}
\title{%
\vskip-3cm{\baselineskip14pt
\centerline{\small DESY~13-239,~~HU-EP-13/76\hfill December 2013}}
\vskip1.5cm
Application of chiral resonance
Lagrangian theories to the muon $g-2$%
%\thanks{Talk at Matter to the Deepest, Ustro\'n, Poland, September 4, 2013}%
}
\author{Fred Jegerlehner
\address{Humboldt-Universit\"at zu Berlin, Institut f\"ur Physik,
       Newtonstrasse 15,\\ D-12489 Berlin, Germany\\
Deutsches Elektronen-Synchrotron (DESY), Platanenallee 6,\\ D-15738 Zeuthen, Germany}
}
\maketitle
\begin{abstract}
%Low lying resonances like the spin 1 resonances $\rho$, $\omega$ and
%$\phi$ are non-perturbative objects in QCD as well as in chiral
%perturbation theory, but play a central role in low energy hadron
%physics. Any halfway realistic phenomenological description has to
%incorporate the low lying particle spectrum in an effective resonance
%Lagrangian. The aim is to single out an optimal phenomenological
%description by a global fit of all available data. As a first step we
%restrict ourselves to describe all two-body reactions together with
%the $\pi^+\pi^-\pi^0$ channel. Remarkably, one indeed can get good
%quality global fits, as far as there are no substantial clashes
%between data from different experiments, as observed for example
%between the KLOE and BaBar $\epm \to \pipi$ cross sections obtained
%with the initial state radiation method.
We think that phenomenological resonance Lagrangian models,
constrained by global fits from low energy hadron reaction data, can
help to improve muon $g-2$ predictions. The main issue are those
contributions which cannot be calculated by perturbative means: the
hadronic vacuum polarization (HVP) effects and the hadronic
light--by--light (HLbL) scattering contribution.  I review recent
progress in the evaluation of the HVP contribution within the broken
Hidden Local Symmetry (HLS) framework, worked out in collaboration with
M.~Benayoun, P.~David and L.~DelBuono.~\cite{Benayoun:2011mm}.
Our HLS driven estimate reads  $a_\mu^{\rm LO~had}=(688.60\pm4.24)\power{-10}$
and we find $a_\mu^{\rm the}= (11 659 177.65 \pm 5.76)\power{-10}$.
\end{abstract}

\section{Effective field theory: the Resonance Lagrangian Approach}
The Resonance Lagrangian Approach (RLA) provides an extension of low
energy effective QCD as represented by Chiral Perturbation Theory
(ChPT) to energies up to about 1 GeV. Principles to be included are
the chiral structure of QCD, the vector-meson dominance model and
electromagnetic gauge invariance. Specifically, we will consider the
HLS version, which is considered to be
equivalent to alternative variants after implementing appropriate high
energy asymptotic conditions. ChPT is the systematic and unambiguous
approach to low energy effective QCD given by spontaneously broken
chiral symmetry $SU(3)\otimes SU(3)$, with the pseudoscalars as
Nambu-Goldstone bosons, together with a systematic expansion in low
momenta and chiral symmetry breaking (SB) effects by the light quark
masses, $m_q\,,\: q=u,d,s$.  The limitation of ChPT is the fact
that it ceases to converge for energies above about $400~\mv$, in
particular it lacks to describe physics involving the vector
resonances $\rho,\omega$ and $\phi$.

The Vector-meson Dominance Model (VDM) is the effective theory
implementing the direct coupling of the neutral spin 1 vector resonances
$\rho,\omega,\phi$ etc. to the photon. Such direct couplings are a
consequence of the fact that the neutral spin 1 resonances like the
$\rho^0$ are composed of charged quarks.
%In the quark model photons
%obviously couple to hadrons via the charged quarks
%\centerline{\includegraphics[height=1.3cm]{VMDlinkc}\raisebox{3ex}{$\epo$}}
The effect is well modeled by the VDM Lagrangian
\mbo{\cL_{\gamma\rho}=\frac{e}{2g_\rho}\,\rho_{\mu\nu}F^{\mu\nu}}
\mbo{\;{\rm \ or \ } =-\frac{eM_\rho^2}{g_\rho}\,\rho_\mu A^\mu},
which has to be implement in low energy effective QCD in a way which
is consistent with the chiral structure of QCD.

The construction of the HLS model may be outlined as follows:
like in ChPT the basic fields are the unitary matrix
fields $\xi_{L,R}=\exp \left[\pm \I\,P/f_\pi\right]$, where
$P=P_8+P_0$
is the $SU(3)$ matrix of pseudoscalar fields, with $P_0$
and $P_8$ the basic singlet and octet fields, respectively.
The pseudoscalar field matrix $P$ is represented by
\Ba
P_8 &=&\frac{1}{\sqrt{2}}
  \left( \begin{array}{ccc}
          \displaystyle  \frac{1}{\sqrt{2}}\pi_3+\frac{1}{\sqrt{6}}\eta_8
            &\displaystyle \pi^+ & \displaystyle  K^+ \\
            \displaystyle \pi^-
& \displaystyle -\frac{1}{\sqrt{2}}\pi_3+\frac{1}{\sqrt{6}}\eta_8 &
\displaystyle K^0 \\
            \displaystyle K^- &  \displaystyle \overline{K}^0
& \displaystyle -\sqrt{\frac{2}{3}}\eta_8
                                             \end{array}
  \right),\\
P_0&=&\frac{1}{\sqrt{6}}{\rm diag}(\eta^0,\eta^0,\eta^0)\semis
(\pi_3,\eta_8,\eta_0)\Leftrightarrow (\pi^0,\eta,\eta')\epo
\Ea
The HLS ansatz is an extension of the ChPT \textit{non-linear sigma
model} to a \textit{non-linear chiral Lagrangian} [\mbo{\Tr\,\partial_\mu \xi^+
\partial^\mu \xi}] based on the symmetry pattern
$G_{\rm global}/H_{\rm local}$, where $G=SU(3)_L \otimes SU(3)_R$ is
the chiral group of QCD and $H=SU(3)_V$ the vector subgroup. The hidden
local $SU(3)_V$ requires the spin 1 vector meson fields, represented
by the $SU(3)$ matrix field $V_\mu$, to be gauge fields.  The
needed covariant derivative reads
$D_\mu=\partial_\mu-i\,g\,V_\mu$, and allows to include the couplings to the
electroweak gauge fields $A_\mu$, $Z_\mu$ and $W^\pm_\mu$ in a natural
way. The vector field matrix is usually written as\\[-5mm]
\Ba
V=\frac{1}{\sqrt{2}}
  \left( \begin{array}{ccc}
   \displaystyle (\rho^I+\omega^I)/\sqrt{2}  & \displaystyle \rho^+             &  \displaystyle K^{*+} \\[0.5cm]
    \displaystyle  \rho^-    & \displaystyle  (-\rho^I+\omega^I)/\sqrt{2}    &  \displaystyle  K^{*0} \\[0.5cm]
      \displaystyle        K^{*-}           & \displaystyle  \overline{K}^{*0}  &  \displaystyle  \phi^I
         \end{array}
  \right)_.
\Ea
The unbroken HLS Lagrangian is then given by
\Ba
\cL_{\rm HLS}=\cL_A+\cL_V\,\semis\cL_{A/V}=-\frac{f_\pi^2}{4}\,\Tr \left[L\pm R\right]^2\,,
\Ea
where $L=\left[D_\mu \xi_L \right]\,\xi_L^+$ and $R=\left[D_\mu \xi_R
\right]\,\xi_R^+$. The covariant derivatives read
\Ba
\left \{
\begin{array}{ccc}
D_\mu \xi_L  = \displaystyle  \pa_\mu \xi_L -i g V_\mu \xi_L +i \xi_L {\cal L}_\mu\\[0.5cm]
D_\mu \xi_R  = \displaystyle  \pa_\mu \xi_R -i g V_\mu \xi_R +i \xi_R {\cal R}_\mu
 \end{array}
 \right._,
\Ea
with known couplings to the Standard Model (SM) gauge bosons
\Ba
\left \{
\begin{array}{l}
{\cal L}_\mu =   \displaystyle  e Q A_\mu + \frac{g_2}{\cos{\theta_W}} (T_z -\sin^2{\theta_W})Z_\mu
+\frac{g_2}{\sqrt{2}} (W^+_\mu T_+ + W^-_\mu T_-)\\[0.5cm]
{\cal R}_\mu =   \displaystyle e Q A_\mu - \frac{g_2}{\cos{\theta_W}} \sin^2{\theta_W} Z_{\mu\;\epo}
 \end{array}
 \right.
\Ea
Like in the electroweak SM, masses of the spin 1 bosons may be
generated by the Higgs-Kibble mechanism if one starts in place of the
non-linear $\sigma$-model with the Gell-Mann--Levy linear
$\sigma$-model by a shift of the $\sigma$-field.

In fact the global chiral symmetry $G_{\rm global}$ is well known not
to be realized as an exact symmetry in nature, which implies that the
ideal HLS symmetry evidently is not a symmetry of nature either. It
evidently has to be broken appropriately in order to provide a
realistic low energy effective theory mimicking low energy effective
QCD. Corresponding to the strength of the breaking, usually, this has
is in two steps, breaking of $SU(3)$ in a first step and breaking
the isospin $SU(2)$ subgroup in a second step. Unlike in ChPT
(perturbed non-linear $\sigma$--model) where one is performing a
systematic low energy expansion, expanding in low momenta and the
quark masses, here we introduce symmetry breaking as phenomenological
parameters to be fixed from appropriate data, since a systematic low
energy expansion a l\'a ChPT ceases to converge at energies above
about 400 MeV, while we attempt to model phenomenology up to including
the $\phi$ resonance.

The broken HLS Lagrangian (BHLS) is then given by (see~\cite{Benayoun:2011mm})
\Ba
\cL_{\rm
BHLS}=\cL'_A+\cL'_V+\cL_{\rm 't Hooft}\,\semis \cL'_{A/V}=-\frac{f_\pi^2}{4}\,\Tr \left\{ \left[L\pm R\right]\,X_{A/V}\right\}^2\;,
\Ea
with 6 phenomenological chiral SB parameters. The phenomenological SB
pattern suggests $ X_I ={\rm diag}(q_I,y_I,z_I)\,,\:
|q_I-1|,|y_I-1|\ll |z_I-1|\,,\: I=V,A\,.$ There is also the parity odd
anomalous sector, which is needed to account for reactions like
$\gamma^* \to \pi^0 \gamma$ and $\gamma^* \to \pi^+\pi^-\pi^0$ among others.

We note that this BHLS model would be a reliable low
energy effective theory if the QCD scale $\Lambda_{\rm QCD}$ would be
large relative to the scale of about 1 GeV up to which we want to
apply the model, which in reality is not the case. Nevertheless, as a
phenomenological model applied to low multiplicity hadronic processes
(specified below) it seems to work pretty well, as we have
demonstrated by a global fit of the available data in
Ref.~\cite{Benayoun:2011mm}.  The major achievement is a simultaneous
consistent fit of the $\eepp$ data from
CMD-2~\cite{Akhmetshin:2006bx}, SND~\cite{Achasov:2006vp},
KLOE~\cite{Ambrosino:2008aa} and BaBar~\cite{Aubert:2009ad}, and the
$\tau\to \pi^-\pi^0 \nu_\tau$ decay spectral functions by
ALEPH~\cite{Schael:2005am}, OPAL~\cite{Ackerstaff:1998yj},
CLEO~\cite{Anderson:1999ui} and Belle~\cite{Fujikawa:2008ma}.  The
$\epm \to\pi^-\pi^+$ channel gives the dominant hadronic contribution
to the muon $g-2$. Isospin symmetry $\pi^-\pi^0\Leftrightarrow
\pi^-\pi^+$ allows one to include existing high quality $\tau$-data as
advocated long time ago in~\cite{Alemany:1997tn}.

We note that as long as higher order corrections are restricted to
the mandatory pion- and Kaon-loop effects in the vector boson
self-energies, renormalizability is not an issue. These contributions
behave as in a strictly renormalizable theory and correspond to a
reparametrization only.

\section{$\rho^0-\gamma$ mixing solving the \mbo{\tau} vs.
\mbo{\epm} puzzle}
\begin{figure}[h]
\centering
\includegraphics[height=1.3cm]{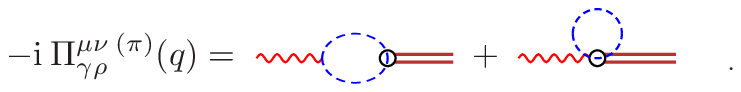}\\
\caption{Irreducible self-energy contribution at one-loop}
\label{fig:rhogammSE}
\end{figure}

A minimal subset of any resonance Lagrangian is given by the
VDM + scalar QED part which describes the leading interaction between the $\rho$ the pions
and the photon. In order to account for the decay of the $\rho$, one
has to include self-energy effects, which also affect $\rho-\gamma$ mixing
via pion-loops shown in Fig.~\ref{fig:rhogammSE}.
Most previous calculations,
considered the mixing term to be a constant, and were missing a substantial quantum
interference effect. The properly normalized
pion form factor, in our approach, has the from
\Ba
F_\pi(s) =  \left[e^2\,D_{\gamma\gamma}+ e\,(g_{\rho\pi\pi}-g_{\rho ee})\,D_{\gamma \rho}-
g_{\rho ee} g_{\rho\pi\pi} \,D_{\rho\rho}\right]/\left[e^2\,D_{\gamma\gamma}\right]\;,
\Ea
with propagators including the pion loop effects,
\begin{figure}[!h]
\centering
\includegraphics[height=5cm]{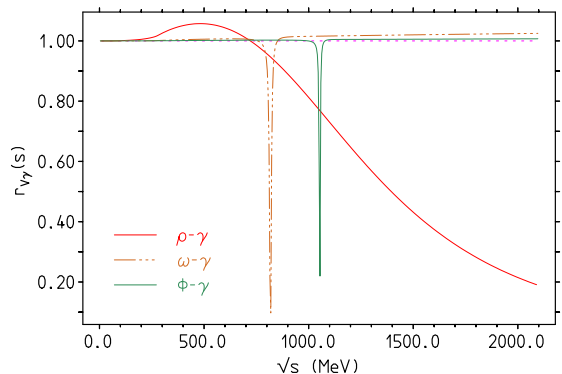}
\caption{Ratio of the full $|F_\pi(s)|^2$
in units of the same quantity omitting the mixing term (full
line). Also shown is the same mechanism scaled up by the branching fraction $\Gamma_V/\Gamma(V\to\pi\pi)$
for $V=\omega$ and $\phi$. In the $\pi\pi$ channel the
effects for resonances $V\neq \rho$ are tiny if not very close to resonance.}
\label{fig:mixingcorr}
\end{figure}
with typical couplings
$ g_{\rho\pi\pi\,\mathrm{bare}} = 5.8935$,
$ g_{\rho\pi\pi\, \mathrm{ren}} = 6.1559$, $ g_{\rho ee} =  0.018149$,
$ x=g_{\rho\pi\pi}/g_\rho=   1.15128\;,$ fixed from the (partial) widths
$${ g_{\rho\pi\pi}}=\sqrt{{48\,\pi}\, \Gamma_\rho/(\beta_\rho^3\,M_\rho)}
\semis {  g_{\rho ee}}=\sqrt{12\pi\,\Gamma_{\rho
ee}/M_\rho}\;.$$ The effect of taking into account or not the
$\gamma-\rho^0$ mixing is illustrated in Fig.~\ref{fig:mixingcorr}.
The $\gamma-\rho$ interference is crucial when relating charged
current $\tau$-data to $\epm$-data. Including known isospin
breaking (IB) corrections $\varv_0(s)=R_{\rm IB}(s)\,\varv_-(s)$ a large discrepancy
[$\sim$ 10\%] persisted~\cite{Davier:2009ag}, which was known as the
$\tau$ vs. $\epm$ puzzle
since~\cite{Davier:2002dy}. In~\cite{Jegerlehner:2011ti} it has been
shown that the $\gamma-\rho$ mixing active in the $\eepp$ channel is
responsible for the discrepancy, i.e. $\tau$-data have to be
corrected as $\varv_0(s)=r_{\rho\gamma}(s)\,R_{\rm
IB}(s)\,\varv_-(s)$, before they can be used as representing an
equivalent I=1 $\eepp$ data sample (see
also~\cite{Benayoun:2007cu,Benayoun:2009im}).
Note that what goes into \mbo{a_\mu} directly are the \mbo{\epm}-data.
\begin{figure}[!h]
\centering
\includegraphics[height=5.4cm]{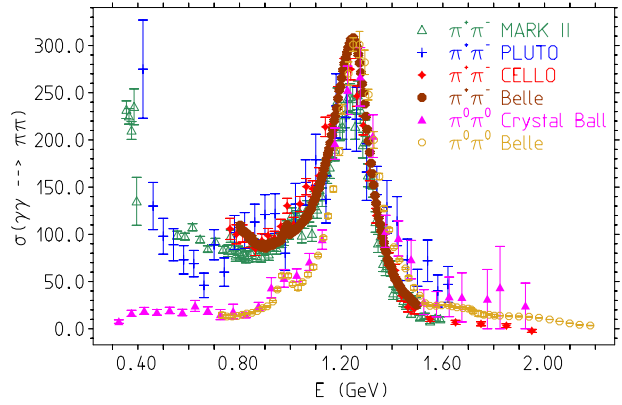}
\caption{How photons couple to pions? This is obviously probed in
reactions like $ \gamma \gamma \to \pi^+\pi^-,\pi^0\pi^0$. Data infer
that below about 1 GeV photons couple to pions as point-like objects
(i.e. to the charged ones overwhelmingly). At higher energies the
photons see the quarks exclusively and form the prominent tensor
resonance $ f_2(1270)$. The $ \pi^0\pi^0$ cross section shown has been
multiplied by the isospin symmetry factor 2, by which it is reduced
in reality.}
\label{fig:ggpipiall}
\end{figure}
Best ``proof'' of the required $\rho-\gamma$ correction profile is the
ALEPH vs. BaBar fit shown in Fig.~1 of~\cite{Davier:2009zi}. Applying
the correction to the $\tau$ spectra (see Fig.~8
in~\cite{Jegerlehner:2011ti}) implies a universal shift down by
$\delta \amuh [\rho\gamma]\simeq (-5.1\pm0.5)\power{-10}$ of the
contribution to the muon $g-2$. This shift brings into agreement the
$\tau$ inclusive estimates with the $\epm$ based ones. Is our model,
treating pions as point-like objects, viable? A good ``answer'' to
this question may be obtained by looking at the $\pi\pi$ production in
$\gamma\gamma$ fusion. Fig.~\ref{fig:ggpipiall} shows: at the strong
tensor meson resonance \mbo{f_2(1270)} in the \mbo{\pi\pi} channel,
photons directly probe the quarks! However, in the
region of our interest photons see pions (below about 1 GeV). We apply the
sQED model up to 0.975 GeV (relevant for
\mbo{a_\mu}), which should be rather reliable.
Switching off the electromagnetic interaction of pions, is definitely
not a realistic approximation in trying to describe what is observed
in the \mbo{\epm \to \ppm} channel.
\section{Global fit of BHLS parameters and prediction
of \mbo{F_\pi(s)}} The simple model just considered illustrates one of
the main quantum interference effects in the isospin sector, the
$\gamma-\rho^0$ mixing. A more complete effective theory must include
the $\rho^0-\omega$ mixing, as well as the strangeness sector, with
the Kaons as additional pseudo Nambu-Goldstone bosons, including the
$\eta$ and the $\eta'$, and the mixing with the $\phi$. This is
implemented in the BHLS model introduced before. Self-energy
corrections for $\rho,\:\omega,\:\phi$ and $\gamma$ now include
Kaon-loops as well. In addition, parity odd sector contributions like
$\pi^0\to\gamma\gamma$ and $\gamma \to \pi^+\pi^-\pi^0$ must be
included. At present there are 45 different data sets (6 annihilation
channels and 10 partial width decays) available below \mbo{E_0 =
1.05~\gv} (just above the
\mbo{\phi}), and we use them to constrain the BHLS Lagrangian
couplings. The method is able to reduce uncertainties in $g-2$
predictions by using indirect constraints on the Lagrangian
parameters.

The main goal is to single out a representative effective resonance
Lagrangian by the global fit. The constrained model is expected to
help in improving model calculations of hadronic light-by-light
scattering.  The new muon g-2 experiments planned at Fermilab and
J-PARC, supposed to start in about 2-3 years, are expected to reduce
experimental errors by a factor 4. On the theory side this requires a
comparable improvement of the HVP and HLbL contributions.

\noi
The effective theory predicts the cross sections
\mbo{\ppm,~\pi^0\gamma,~\eta\gamma,~\eta'\gamma,~\pi^0\pi^+\pi^-,}
\mbo{~K^+K^-,~K^0\bar{K}^0} which account 83.4\% of the HVP
contribution to the muon $g-2$. Contributions from the missing channels
\mbo{4\pi,5\pi,6\pi,\eta\pi\pi,\omega\pi} and from higher energies we
evaluate using data directly and pQCD in the perturbative region and
in the tail. The resulting BHLS prediction for \mbo{a_\mu^{\rm
LO,had}} allows us to get a BHLS driven SM prediction for $a_\mu$ (see
Table 1). Our
favored evaluation based on selected data yields $a_\mu^{\rm
LO~had}=(681.23\pm4.51)\power{-10}$ and a prediction $a_\mu^{\rm the}=
(11 659 170.28 \pm 5.96)\power{-10}$ and $\Delta a_\mu= a_\mu^{\rm
exp}-a_\mu^{\rm the}=(38.52 \pm 5.96_{\rm the}
\pm 6.3_{\rm exp})\power{-10}\epo$
The associated fit probability is 94\% and the significance for
$\Delta a_\mu$ is $4.4\sigma$. Including all data, applying appropriate
rewighting in case of inconsistencies\footnote{The required
rewighting concerns the $e^+e^- \to \pi^+\pi^-\pi^0$ data in the vicinity
of the $\phi$, as well as the KLOE08 and the BaBar $e^+e^- \to
\pi^+\pi^-$ data sets.}, we find $ a_\mu^{\rm LO~had}=(688.60\pm4.24)\power{-10}$
such that $a_\mu^{\rm the}= (11 659 177.65 \pm 5.76)\power{-10}$ and
$\Delta a_\mu= a_\mu^{\rm exp}-a_\mu^{\rm the}=
 (31.25 \pm 5.76_{\rm the} \pm 6.3_{\rm exp})\power{-10}\epo$
The associated fit probability is 76\%
and the significance for $\Delta a_\mu$ is $3.7\sigma$.
\begin{figure}
\centering
\includegraphics[height=8cm]{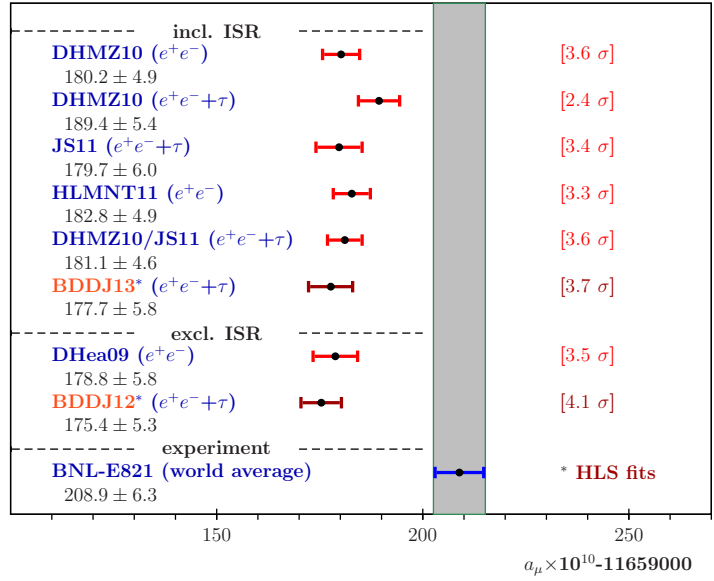}
\caption{Comparison with other Results. Note: results depend on which
value is taken for HLbL. JS11 and BDDJ13 includes $116(39)\power{-11}$ [JN~\cite{JN09}],
DHea09, DHMZ10, HLMNT11 and BDDJ12 use $105(26)\power{-11}$
[PdRV~\cite{Prades:2009tw}].}
\label{fig:compare}
\end{figure}
The comparison of our global fit result with other results from
DHMZ10~\cite{Davier:2009zi,Davier:2010nc},
JS11~\cite{Jegerlehner:2011ti}, DHea09~\cite{Davier:2009ag},
HLMNT11~\cite{Hagiwara:2011af} is shown in Fig.~\ref{fig:compare}.  We
get somewhat lower central values than results obtained by direct
integration of the data, but all results agree well within
1$\sigma$. Our fits, which include the $\tau$ data, exhibit the best
fit probability for KLOE10 results, while there is some tension
showing up in case of the BaBar $\pi\pi$ data. Our analysis has been
criticized lately in Ref.~\cite{Davier:2013rla} but what is shown in
that reply is that BaBar~\cite{Aubert:2009ad} and KLOE data are not
quite compatible within the given experimental errors. A different
issue is the comparison between BaBar and $\tau$ spectral
data. Contrary to claims in~\cite{Davier:2013rla} the sizable
$\gamma-\rho^0$ mixing effect has \textbf{not} been taken into account
and one should see a substantial shift which, however, is found to be
absent in the comparison between Belle $\tau$ data and the BaBar
$\epm$ data (see Fig. 1 in~\cite{Davier:2009zi}).

A comparison between theory and experiment~\cite{Bennett:2006fi} is
given in Tab.~\ref{tab:theory} (see
also~\cite{Miller:2012opa}). Theory results shown are updates from
Ref.~\cite{JN09} using results on improved 4-loop and the new
5-loop QED corrections~\cite{Aoyama:2012wk}, improved lepton mass
ratios~\cite{Mohr:2012tt} and using the new Higgs mass value from
ATLAS and CMS in the evaluation of the weak
corrections~\cite{Gnendiger:2013pva}.
\begin{table}[t]
%\vspace*{-0mm}
\begin{center}
\caption{Standard model theory and experiment
comparison [in units $10^{-11}$].
%comparison.
\label{tab:theory}}
\centering
\begin{tabular}{lr@{.}lr@{.}lcc}
&\ttc{}~&\ttc{}&&\\[-3mm]
\hline
&\ttc{}~&\ttc{}&&\\[-3mm]
%Contribution & \multicolumn{2}{c}{Value~$\times 10^{10}$} & \multicolumn{2}{c}{Error~$\times 10^{10}$} & Reference \\
Contribution & \multicolumn{2}{c}{Value} & \multicolumn{2}{c}{Error} \\
&\ttc{}~&\ttc{}\\[-3mm]
\hline\noalign{\smallskip}
QED incl. 4-loops + 5-loops & 116\,584\,718&85 & 0&04 \\
Leading hadronic vacuum polarization & 6\,886&0 & ~~42&4 \\
Subleading hadronic vacuum polarization & -98&32 & 0&82 \\
Hadronic light--by--light &  116&0 & 39&0 \\
Weak incl. 2-loops & 154&0 & 1&0 \\
&\ttc{}~&\ttc{}&\\
Theory        & 116\,591\,776&5 & 57&6  \\
Experiment & 116\,592\,089&0 & 63&0 \\
Exp. - The. ~~  3.7 standard deviations & 312&5 & 85&4 \\ \noalign{\smallskip}\hline
\end{tabular}
\vspace*{-4mm}
\end{center}
% updated QED, hadronic VP, weak July 2012
\end{table}

\section{Lessons and Outlook}

Effective field theory is the only way to understand relationships
between different channels, like \mbo{\epm}--annihilation
cross-sections and \mbo{\tau}--decay spectra. Global fit strategies
allow to single out variants of effective resonance Lagrangian
models. Models for individual channels can parametrize data, but do
not allow to understand them and their relation to other channels.  We
get perfect fits for \mbo{|F_\pi(s)|^2} up to just above the
\mbo{\phi} without higher \mbo{\rho}'s \mbo{\rho^{\raisebox{-2ex}{$\,'$}},\rho^{\raisebox{-2ex}{$\,''$}}}, which
seem to be mandatory in Gounaris-Sakurai type fits.
\mbo{\tau} data in our approach play a special role, because they are much simpler
than the \mbo{\epm} data, which exhibit intricate
\mbo{\gamma-\rho^0-\omega-\phi} mixing effects.

RLA type analyses provide analytic shapes for amplitudes, and such
``physical shape information'' is favorable over ad hoc data
interpolations, the simplest being the trapezoidal rule, which is known to be
problematic when data are sparse or strongly energy dependent.

Limitations of the RLA are the large couplings which make systematic
higher order improved analyses problematic. As illustrated by
Fig.~\ref{fig:ggpipiall}, considering pions and Kaons to be point-like
may be not too bad an approximation, in the range we are applying the
model. Also, we consider our analysis as a starting point to be
confronted with other RLA versions and implementations and with what
happens if one tries to include higher order effects.

\noindent
{\bf Acknowledgments} \\ Many thanks to the organizers for the
invitation and support to the 2013 ``Matter to the Deepest'' International
Conference at Ustro\'n, Poland, and for giving me the opportunity to present
this talk.


\begin{thebibliography}{99.}

\bibitem{Benayoun:2011mm}
  M.~Benayoun, P.~David, L.~DelBuono, F.~Jegerlehner,
  %``Upgraded Breaking Of The HLS Model: A Full Solution to the $\tau^-e^+e^-$ and $\phi$ Decay Issues And Its Consequences On g-2 VMD Estimates,''
  Eur.\ Phys.\ J.\ C {\bf 72} (2012) 1848;
%  [arXiv:1106.1315 [hep-ph]].
%  %%CITATION = ARXIV:1106.1315;%%
%  %30 citations counted in INSPIRE as of 17 Oct 2013
%
%\bibitem{Benayoun:2012wc}
%  M.~Benayoun, P.~David, L.~DelBuono, F.~Jegerlehner,
%  %``An Update of the HLS Estimate of the Muon g-2,''
  Eur.\ Phys.\ J.\ C {\bf 73} (2013) 2453.
%  [arXiv:1210.7184 [hep-ph]].
  %%CITATION = ARXIV:1210.7184;%%
  %20 citations counted in INSPIRE as of 17 Oct 2013

\bibitem{Akhmetshin:2006bx}
  R.~R.~Akhmetshin et al.  [CMD-2 Collaboration],
  %``Reanalysis of hadronic cross-section measurements at CMD-2,''
  Phys.\ Lett.\ B {\bf 578} (2004) 285;
  %``High-statistics measurement of the pion form factor in the rho-meson energy range with the CMD-2 detector,''
  Phys.\ Lett.\ B {\bf 648} (2007) 28.
%  %%CITATION = HEP-EX/0610021;%%

\bibitem{Achasov:2006vp}
  M.~N.~Achasov et al,
%  M.~N.~Achasov, K.~I.~Beloborodov, A.~V.~Berdyugin, A.~G.~Bogdanchikov, A.~V.~Bozhenok, A.~D.~Bukin, D.~A.~Bukin and T.~V.~Dimova {\it et al.},
  %``Update of the e+ e- ---> pi+ pi- cross-section measured by SND detector in the energy region 400-MeV < s**(1/2) < file1000-MeV,''
  J.\ Exp.\ Theor.\ Phys.\  {\bf 103} (2006) 380
   [Zh.\ Eksp.\ Teor.\ Fiz.\  {\bf 130} (2006) 437].
  %%CITATION = HEP-EX/0605013;%%
  %96 citations counted in INSPIRE as of 17 Oct 2013

\bibitem{Ambrosino:2008aa}
  F.~Ambrosino et al.  [KLOE Collaboration],
  %``Measurement of sigma(e + e- ---> pi+ pi- gamma(gamma) and the dipion contribution to the muon anomaly with the KLOE detector,''
  Phys.\ Lett.\ B {\bf 670} (2009) 285;
%  [arXiv:0809.3950 [hep-ex]].
%  %%CITATION = ARXIV:0809.3950;%%
%  %80 citations counted in INSPIRE as of 17 Oct 2013
%\bibitem{Ambrosino:2010bv}
%  F.~Ambrosino et al.  [KLOE Collaboration],
  %``Measurement of sigma(e+ e- -> pi+ pi-) from threshold to 0.85 GeV^2 using Initial State Radiation with the KLOE detector,''
  Phys.\ Lett.\ B {\bf 700} (2011) 102;
%  [arXiv:1006.5313 [hep-ex]].
  %%CITATION = ARXIV:1006.5313;%%
  %54 citations counted in INSPIRE as of 17 Oct 2013
%\bibitem{Babusci:2012rp}
  D.~Babusci et al.  [KLOE Collaboration],
  %``Precision measurement of $\sigma(e^+e^-\rightarrow \pi^+\pi^-\gamma)/ \sigma(e^+e^-\rightarrow \mu^+\mu^-\gamma)$ and determination of the $\pi^+\pi^-$ contribution to the muon anomaly with the KLOE detector,''
  Phys.\ Lett.\ B {\bf 720} (2013) 336.
%  [arXiv:1212.4524 [hep-ex]].
  %%CITATION = ARXIV:1212.4524;%%
  %6 citations counted in INSPIRE as of 18 Oct 2013

\bibitem{Aubert:2009ad}
  B.~Aubert et al. [BaBar Collaboration],
  %``Precise measurement of the e+ e- ---> pi+ pi- (gamma) cross section with the Initial State Radiation method at BABAR,''
  Phys.\ Rev.\ Lett.\  {\bf 103} (2009) 231801;
%  %%CITATION = ARXIV:0908.3589;%%
%  %105 citations counted in INSPIRE as of 17 Oct 2013
%\bibitem{Lees:2012cj}
  J.~P.~Lees et al. [BaBar Collaboration],
  %``Precise Measurement of the $e^+ e^- \to \pi^+\pi^- (\gamma)$ Cross Section with the Initial-State Radiation Method at BABAR,''
  Phys.\ Rev.\ D {\bf 86} (2012) 032013.
  %%CITATION = ARXIV:1205.2228;%%
  %14 citations counted in INSPIRE as of 17 Oct 2013

\bibitem{Schael:2005am}
  S.~Schael et al.  [ALEPH Collaboration],
  %``Branching ratios and spectral functions of tau decays: Final ALEPH measurements and physics implications,''
  Phys.\ Rept.\  {\bf 421} (2005) 191.
  %%CITATION = HEP-EX/0506072;%%
  %215 citations counted in INSPIRE as of 17 Oct 2013

\bibitem{Ackerstaff:1998yj}
  K.~Ackerstaff et al.  [OPAL Collaboration],
  %``Measurement of the strong coupling constant alpha(s) and the vector and axial vector spectral functions in hadronic tau decays,''
  Eur.\ Phys.\ J.\ C {\bf 7} (1999) 571.
  %%CITATION = HEP-EX/9808019;%%
  %300 citations counted in INSPIRE as of 17 Oct 2013

\bibitem{Anderson:1999ui}
  S.~Anderson et al.  [CLEO Collaboration],
  %``Hadronic structure in the decay tau- ---> pi- pi0 neutrino(tau),''
  Phys.\ Rev.\ D {\bf 61} (2000) 112002.
  %%CITATION = HEP-EX/9910046;%%
  %160 citations counted in INSPIRE as of 17 Oct 2013

\bibitem{Fujikawa:2008ma}
  M.~Fujikawa et al.  [Belle Collaboration],
  %``High-Statistics Study of the tau- ---> pi- pi0 nu(tau) Decay,''
  Phys.\ Rev.\ D {\bf 78} (2008) 072006.
  %%CITATION = ARXIV:0805.3773;%%
  %80 citations counted in INSPIRE as of 17 Oct 2013

\bibitem{Alemany:1997tn}
  R.~Alemany, M.~Davier, A.~H\"ocker,
  %``Improved determination of the hadronic contribution to the muon (g-2) and to alpha (M(z)) using new data from hadronic tau decays,''
  Eur.\ Phys.\ J.\ C {\bf 2} (1998) 123.
  %%CITATION = HEP-PH/9703220;%%
  %270 citations counted in INSPIRE as of 17 Oct 2013

\bibitem{Davier:2009ag}
  M.~Davier et al,
%  M.~Davier, A.~Hoecker, G.~Lopez Castro, B.~Malaescu, X.~H.~Mo, G.~Toledo Sanchez, P.~Wang and C.~Z.~Yuan {\it et al.},
  %``The Discrepancy Between tau and e+e- Spectral Functions Revisited and the Consequences for the Muon Magnetic Anomaly,''
  Eur.\ Phys.\ J.\ C {\bf 66} (2010) 127.
%  [arXiv:0906.5443 [hep-ph]].
  %%CITATION = ARXIV:0906.5443;%%
  %88 citations counted in INSPIRE as of 17 Oct 2013

\bibitem{Davier:2002dy}
  M.~Davier, S.~Eidelman, A.~H\"ocker, Z.~Zhang,
  %``Confronting spectral functions from e+ e- annihilation and tau decays: Consequences for the muon magnetic moment,''
  Eur.\ Phys.\ J.\ C {\bf 27} (2003) 497.
%  [hep-ph/0208177].
  %%CITATION = HEP-PH/0208177;%%
  %328 citations counted in INSPIRE as of 18 Oct 2013

\bibitem{Jegerlehner:2011ti}
  F.~Jegerlehner, R.~Szafron,
  %``$\rho^0 - \gamma$ mixing in the neutral channel pion form factor $F_{\pi}^{e}$ and its role in comparing $e^+ e^-$ with $\tau$ spectral functions,''
  Eur.\ Phys.\ J.\ C {\bf 71} (2011) 1632.
%  [arXiv:1101.2872 [hep-ph]].
  %%CITATION = ARXIV:1101.2872;%%
  %62 citations counted in INSPIRE as of 18 Oct 2013
\bibitem{Benayoun:2007cu}
  M.~Benayoun et al,
%P.~David, L.~DelBuono, O.~Leitner, H.~B.~O'Connell,
  %``The Dipion Mass Spectrum In e+e- Annihilation and tau Decay: A Dynamical
  %(rho0, omega, phi) Mixing Approach,''
  Eur.\ Phys.\ J.\  C {\bf 55} (2008) 199.
%  [arXiv:0711.4482 [hep-ph]].
  %%CITATION = EPHJA,C55,199;%%

\bibitem{Benayoun:2009im}
  M.~Benayoun, P.~David, L.~DelBuono, O.~Leitner,
  %``A Global Treatment Of VMD Physics Up To The phi: I. e+ e- Annihilations, Anomalies And Vector Meson Partial Widths,''
  Eur.\ Phys.\ J.\ C {\bf 65} (2010) 211;
%  [arXiv:0907.4047 [hep-ph]].
  %%CITATION = ARXIV:0907.4047;%%
  %27 citations counted in INSPIRE as of 18 Oct 2013
%\bibitem{Benayoun:2009fz}
%  M.~Benayoun, P.~David, L.~DelBuono and O.~Leitner,
%  %``A Global Treatment Of VMD Physics Up To The phi: II. tau Decay and Hadronic Contributions To g-2,''
  Eur.\ Phys.\ J.\ C {\bf 68} (2010) 355.
%  [arXiv:0907.5603 [hep-ph]].
  %%CITATION = ARXIV:0907.5603;%%
  %13 citations counted in INSPIRE as of 18 Oct 2013

\bibitem{Davier:2009zi}
  M.~Davier et al,
% A.~H\"ocker, B.~Malaescu, C.~Z.~Yuan, Z.~Zhang,
  %``Reevaluation of the hadronic contribution to the muon magnetic anomaly using new e+ e- ---> pi+ pi- cross section data from BABAR,''
  Eur.\ Phys.\ J.\ C {\bf 66} (2010) 1.
%  [arXiv:0908.4300 [hep-ph]].
  %%CITATION = ARXIV:0908.4300;%%
  %135 citations counted in INSPIRE as of 18 Oct 2013

\bibitem{JN09}
  F.~Jegerlehner, A.~Nyffeler,
  %``The Muon g-2,''
  Phys.\ Rept.\  {\bf 477} (2009) 1;
  %%CITATION = PRPLC,477,1;%%
  F.~Jegerlehner,
% \textit{The anomalous magnetic moment of the muon,}
  Springer Tracts Mod.\ Phys.\  {\bf 226} (2008) 1;
  %%CITATION = STPHB,226,1;%%
  %5 citations counted in INSPIRE as of 18 Oct 2013
%\bibitem{Jegerlehner:2007xe}
%  F.~Jegerlehner,
  %``Essentials of the Muon g-2,''
%  Acta Phys.\ Polon.\  B {\bf 38} (2007) 3021.
%  [arXiv:hep-ph/0703125].
  %%CITATION = APPOA,B38,3021;%%
%\bibitem{Jegerlehner:2009zz}
%  F.~Jegerlehner,
  %``Progress in the prediction of g-2 of the muon,''
  Acta Phys.\ Polon.\ B {\bf 40} (2009) 3097.
  %%CITATION = APPOA,B40,3097;%%

\bibitem{Prades:2009tw}
  J.~Prades, E.~de Rafael, A.~Vainshtein,
  %``Hadronic Light-by-Light Scattering Contribution to the Muon Anomalous Magnetic Moment,''
%  (Advanced series on directions in high energy physics. 20)
  [arXiv:0901.0306 [hep-ph]].
  %%CITATION = ARXIV:0901.0306;%%
  %107 citations counted in INSPIRE as of 20 Oct 2013

\bibitem{Davier:2010nc}
  M.~Davier, A.~H\"ocker, B.~Malaescu, Z.~Zhang,
  %``Reevaluation of the Hadronic Contributions to the Muon g-2 and to alpha(MZ),''
  Eur.\ Phys.\ J.\ C {\bf 71} (2011) 1515
   [Erratum-ibid.\ C {\bf 72} (2012) 1874].
%  [arXiv:1010.4180 [hep-ph]].
  %%CITATION = ARXIV:1010.4180;%%
  %242 citations counted in INSPIRE as of 17 Oct 2013

\bibitem{Hagiwara:2011af}
  K.~Hagiwara, R.~Liao, A.~D.~Martin, D.~Nomura, T.~Teubner,
  %``(g-2)_mu and alpha(M_Z^2) re-evaluated using new precise data,''
  J.\ Phys.\ G {\bf 38} (2011) 085003.
%  [arXiv:1105.3149 [hep-ph]].
  %%CITATION = ARXIV:1105.3149;%%
  %160 citations counted in INSPIRE as of 17 Oct 2013

\bibitem{Davier:2013rla}
  M.~Davier, B.~Malaescu,
  %``Comments on "An Update of the HLS Estimate of the Muon g-2"by M.Benayoun {\it et al.}, arXiv:1210.7184v3,''
  arXiv:1306.6374 [hep-ex].
  %%CITATION = ARXIV:1306.6374;%%

\bibitem{Bennett:2006fi}
  G.~W.~Bennett et al.  [Muon G-2 Collaboration],
  %``Final Report of the Muon E821 Anomalous Magnetic Moment Measurement at BNL,''
  Phys.\ Rev.\ D {\bf 73} (2006) 072003.
%  [hep-ex/0602035].
  %%CITATION = HEP-EX/0602035;%%
  %673 citations counted in INSPIRE as of 17 Oct 2013

\bibitem{Miller:2012opa}
  J.~P.~Miller, E.~d.~Rafael, B.~L.~Roberts, D.~St\"ockinger,
  %``Muon (g-2): Experiment and Theory,''
  Ann.\ Rev.\ Nucl.\ Part.\ Sci.\  {\bf 62} (2012) 237.
  %%CITATION = ARNUA,62,237;%%
  %3 citations counted in INSPIRE as of 20 Oct 2013

\bibitem{Aoyama:2012wk}
  T.~Aoyama, M.~Hayakawa, T.~Kinoshita, M.~Nio,
  %``Complete Tenth-Order QED Contribution to the Muon g-2,''
  Phys.\ Rev.\ Lett.\  {\bf 109} (2012) 111808.
%  [arXiv:1205.5370 [hep-ph]].
  %%CITATION = ARXIV:1205.5370;%%
  %43 citations counted in INSPIRE as of 17 Oct 2013

\bibitem{Mohr:2012tt}
  P.~J.~Mohr, B.~N.~Taylor, D.~B.~Newell,
  %``CODATA Recommended Values of the Fundamental Physical Constants: 2010,''
  Rev.\ Mod.\ Phys.\  {\bf 84} (2012) 1527.
%  [arXiv:1203.5425 [physics.atom-ph]].
  %%CITATION = ARXIV:1203.5425;%%
  %56 citations counted in INSPIRE as of 17 Oct 2013

\bibitem{Gnendiger:2013pva}
  C.~Gnendiger, D.~St\"ockinger, H.~St\"ockinger-Kim,
  %``The electroweak contributions to (g-2)_\mu\ after the Higgs boson mass measurement,''
  Phys.\ Rev.\ D {\bf 88} (2013) 053005.
%  [arXiv:1306.5546 [hep-ph]].
  %%CITATION = ARXIV:1306.5546;%%
  %3 citations counted in INSPIRE as of 17 Oct 2013

\end{thebibliography}
\end{document}